# A Critique of ANSI SQL Isolation Levels

| | | |
|---|---|---|
| Hal Berenson | Microsoft Corp. | haroldb@microsoft.com |
| Phil Bernstein | Microsoft Corp. | philbe@microsoft.com |
| Jim Gray | Microsoft Corp. | gray@microsoft.com |
| Jim Melton | Sybase Corp. | jim.melton@sybase.com |
| Elizabeth O'Neil | UMass/Boston | eoneil@cs.umb.edu |
| Patrick O'Neil | UMass/Boston | poneil@cs.umb.edu |

# A Critique of ANSI SQL Isolation Levels

| | | |
|---|---|---|
| Hal Berenson | Microsoft Corp. | haroldb@microsoft.com |
| Phil Bernstein | Microsoft Corp. | philbe@microsoft.com |
| Jim Gray | U.C. Berkeley | gray@crl.com |
| Jim Melton | Sybase Corp. | jim.melton@sybase.com |
| Elizabeth O'Neil | UMass/Boston | eoneil@cs.umb.edu |
| Patrick O'Neil | UMass/Boston | poneil@cs.umb.edu |

**Abstract**: ANSI SQL-92 [MS, ANSI] defines Isolation *Levels* in terms of *phenomena:* Dirty Reads, Non-Repeatable Reads, and Phantoms. This paper shows that these phenomena and the ANSI SQL definitions fail to characterize several popular isolation levels, including the standard locking implementations of the levels. Investigating the ambiguities of the phenomena leads to clearer definitions; in addition new phenomena that better characterize isolation types are introduced. An important multiversion isolation type, *Snapshot Isolation*, is defined.

## 1. Introduction

Running concurrent transactions at different isolation levels allows application designers to trade throughput for correctness. Lower isolation levels increase transaction concurrency but risk showing transactions a fuzzy or incorrect database. Surprisingly, some transactions can execute at the highest isolation level (perfect serializability) while concurrent transactions running at a lower isolation level can access states that are not yet committed or that postdate states the transaction read earlier [GLPT]. Of course, transactions running at lower isolation levels may produce invalid data. Application designers must prevent later transactions running at higher isolation levels from accessing this invalid data and propagating errors.

The ANSI/ISO SQL-92 specifications [MS, ANSI] define four isolation levels: (1) READ UNCOMMITTED, (2) READ COMMITTED, (3) REPEATABLE READ, (4) SERIALIZABLE. These levels are defined with the classical serializability definition, plus three prohibited action subsequences, called *phenomena*: *Dirty Read*, *Non-repeatable Read*, and *Phantom*. The concept of a *phenomenon* is not explicitly defined in the ANSI specifications, but the specifications suggest that phenomena are action subsequences that may lead to anomalous (perhaps non-serializable) behavior. We refer to *anomalies* in what follows when suggesting additions to the set of ANSI phenomena. As shown later, there is a technical distinction between anomalies and phenomena, but this distinction is not crucial for a general understanding.

The ANSI isolation levels are related to the behavior of lock schedulers. Some lock schedulers allow transactions to vary the scope and duration of their lock requests, thus departing from pure two-phase locking. This idea was introduced by [GLPT], which defined *Degrees of Consistency* in three ways: locking, data-flow graphs, and anomalies. Defining isolation levels by phenomena (anomalies) was intended to allow non-lock-based implementations of the SQL standard.

This paper shows a number of weaknesses in the anomaly approach to defining isolation levels. The three ANSI phenomena are ambiguous. Even their broadest interpretations do not exclude anomalous behavior. This leads to some counter-intuitive results. In particular, lock-based isolation levels have different characteristics than their ANSI equivalents. This is disconcerting because commercial database systems typically use locking. Additionally, the ANSI phenomena do not distinguish among several isolation levels popular in commercial systems.

Section 2 introduces basic isolation level terminology. It defines the ANSI SQL and locking isolation levels. Section 3 examines some drawbacks of the ANSI isolation levels and proposes a new phenomenon. Other popular isolation levels are also defined. The various definitions map between ANSI SQL isolation levels and the *degrees of consistency* defined in 1977 in [GLPT]. They also encompass Date's definitions of Cursor Stability and Repeatable Read [DAT]. Discussing the isolation levels in a uniform framework reduces confusion.

Section 4 introduces a multiversion concurrency control mechanism, called *Snapshot Isolation,* that avoids the ANSI SQL phenomena, but is not serializable. Snapshot Isolation is interesting in its own right, since it provides a reduced-isolation level approach that lies between READ COMMITTED and REPEATABLE READ. A new formalism (available in the longer version of this paper [OOBBGM]) connects reduced isolation levels for multiversioned data to the classical single-version locking serializability theory.

Section 5 explores some new anomalies to differentiate the isolation levels introduced in Sections 3 and 4. The extended ANSI SQL phenomena proposed here lack the power to characterize Snapshot isolation and Cursor Stability. Section 6 presents a Summary and Conclusions.

## 2. Isolation Definitions

### 2.1 Serializability Concepts

Transactional and locking concepts are well documented in the literature [BHG, PAP, PON, GR]. The next few paragraphs review the terminology used here.

A *transaction* groups a set of actions that transform the database from one consistent state to another. A *history* models the interleaved execution of a set of transactions as a linear ordering of their actions, such as Reads and Writes (i.e., inserts, updates, and deletes) of specific data items. Two actions in a history are said to *conflict* if they are performed by distinct transactions on the same data item and at least one of is a Write action. Following [EGLT], this definition takes a broad interpretation of "data item": it could be a table row, a page, an entire table, or a message on a queue. Conflicting actions can also occur on a *set* of data items, covered by a *predicate lock,* as well as on a single data item.

A particular history gives rise to a *dependency graph* defining the temporal data flow among transactions. The actions of committed transactions in the history are represented as graph nodes. If action op1 of transaction T1 conflicts with and precedes action op2 of transaction T2 in the history, then the pair <op1, op2> becomes an edge in the dependency graph. Two histories are *equivalent* if they have the same committed transactions and the same dependency graph. A history is *serializable* if it is equivalent to a *serial history* — that is, if it has the same dependency graph (inter-transaction temporal data flow) as some history that executes transactions one at a time in sequence.

### 2.2 ANSI SQL Isolation Levels

ANSI SQL Isolation designers sought a definition that would admit many different implementations, not just locking. They defined isolation with the following three *phenomena:*

**P1 (Dirty Read)**: Transaction T1 modifies a data item. Another transaction T2 then reads that data item before T1 performs a COMMIT or ROLLBACK. If T1 then performs a ROLLBACK, T2 has read a data item that was never committed and so never really existed.

**P2 (Non-repeatable or Fuzzy Read)**: Transaction T1 reads a data item. Another transaction T2 then modifies or deletes that data item and commits. If T1 then attempts to reread the data item, it receives a modified value or discovers that the data item has been deleted.

**P3 (Phantom)**: Transaction T1 reads a set of data items satisfying some <search condition>. Transaction T2 then creates data items that satisfy T1's <search condition> and commits. If T1 then repeats its read with the same <search condition>, it gets a set of data items different from the first read.

None of these phenomena could occur in a serial history. Therefore by the Serializability Theorem they cannot occur in a serializable history [EGLT, BHG Theorem 3.6, GR Section 7.5.8.2, PON Theorem 9.4.2].

Histories consisting of reads, writes, commits, and aborts can be written in a shorthand notation: "w1[x]" means a write by transaction 1 on data item x (which is how a data item is "modified"), and "r2[x]" represents a read of x by transaction 2. Transaction 1 reading and writing a set of records satisfying predicate P is denoted by r1[P] and w1[P] respectively. Transaction 1's commit and abort (ROLLBACK) are written "c1" and "a1", respectively.

Phenomenon P1 might be restated as disallowing the following scenario:

(2.1)   w1[x] . . . r2[x] . . . (a1 and c2 in either order)

The English statement of P1 is ambiguous. It does not actually insist that T1 abort; it simply states that if this happens something unfortunate might occur. Some people reading P1 interpret it to mean:

(2.2)   w1[x]...r2[x]...((c1 or a1) and (c2 or a2) in any order)

Forbidding the (2.2) variant of P1 disallows any history where T1 modifies a data item x, then T2 reads the data item before T1 commits or aborts. It does not insist that T1 aborts or that T2 commits.

Definition (2.2) is a much broader interpretation of P1 than (2.1), since it prohibits all four possible commit-abort pairs by transactions T1 and T2, while (2.1) only prohibits two of the four. Interpreting (2.2) as the meaning of P1 prohibits an execution sequence if something anomalous *might* in the future. We call (2.2) the *broad interpretation* of P1, and (2.1) the *strict interpretation* of P1. Interpretation (2.2) specifies a phenomenon that might lead to an anomaly, while (2.1) specifies an actual anomaly. Denote them as P1 and A1 respectively. Thus:

 **P1**: w1[x]...r2[x]...((c1 or a1) and (c2 or a2) in any order)
 **A1**: w1[x]...r2[x]...(a1 and c2 in any order)

Similarly, the English language phenomena P2 and P3 have strict and broad interpretations, and are denoted P2 and P3 for broad, and A2 and A3 for strict:

 **P2:** r1[x]...w2[x]...((c1 or a1) and (c2 or a2) in any order)
 **A2:** r1[x]...w2[x]...c2...r1[x]...c1
 **P3:** r1[P]...w2[y in P]...((c1 or a1) and (c2 or a2) any order)
 **A3:** r1[P]...w2[y in P]...c2...r1[P]...c1

Section 3 analyzes these alternative interpretations after more conceptual machinery has been developed. It argues that the broad interpretation of the phenomena is required. Note that the English statement of ANSI SQL P3 just prohibits inserts to a predicate, but P3 above intentionally prohibits *any* write (insert, update, delete) affecting a tuple satisfying the predicate once the predicate has been read.

This paper later deals with the concept of a *multi-valued history* (*MV-history* for short — see [BHG], Chapter 5). Without going into details now, multiple versions of a data item may exist at one time in a multi-version system. Any read must be explicit about which version is being read. There have been attempts to relate ANSI Isolation definitions to multi-version systems as well as more common single-version systems of a standard locking scheduler. The English language statements of the phenomena P1, P2, and P3 imply single-version histories. This is how we interpret them in the next section.

ANSI SQL defines four *levels of isolation* by the matrix of Table 1. Each isolation level is characterized by the phenomena that a transaction is forbidden to experience (broad or strict interpretations). However, the ANSI SQL specifications do not define the SERIALIZABLE isolation level solely in terms of these phenomena. Subclause 4.28, "SQL-transactions", in [ANSI] notes that the SERIALIZABLE isolation level must provide what is "commonly known as fully serializable execution." The prominence of the table compared to this extra proviso leads to a common misconception that disallowing the three phenomena implies serializability. Table 1 calls histories that disallow the three phenomena ANOMALY SERIALIZABLE.

The isolation levels are defined by the phenomena they are forbidden to experience. Picking a broad interpretation of a phenomenon excludes a larger set of histories than the strict interpretation. This means we are arguing for more *restrictive* isolation levels (more histories will be disallowed). Section 3 shows that even taking the broad interpretations of P1, P2, and P3, forbidding these phenomena does not guarantee true serializability. It would have been simpler in [ANSI] to drop P3 and just use Subclause 4.28 to define ANSI SERIALIZABLE. Note that Table 1 is not a final result; Table 3 will superseded it.

## 2.3 Locking

Most SQL products use lock-based isolation. Consequently, it is useful to characterize the ANSI SQL isolation levels in terms of locking, although certain problems arise.

Transactions executing under a locking scheduler request *Read* (Share) and *Write* (Exclusive) *locks* on data items or sets of data items they read and write. Two locks by different transactions on the same item *conflict* if at least one of them is a Write lock.

A *Read (*resp*. Write) predicate lock* on a given `<search condition>` is effectively a lock on all data items satisfying the `<search condition>`. This may be an infinite set. It includes data present in the database and also any *phantom data items* not currently in the database but that *would* satisfy the predicate if they were inserted or if current data items were updated to satisfy the `<search condition>`. In SQL terms, a predicate lock covers all tuples that satisfy the predicate and any that an `INSERT`, `UPDATE`, or `DELETE` statement would cause to satisfy the predicate. Two *predicate locks* by different transactions *conflict* if one is a Write lock and if there is a (possibly phantom) data item covered by both locks. An item lock (record lock) is a predicate lock where the predicate names the specific record.

A transaction has *well-formed writes* (*reads*) if it requests a Write (Read) lock on each data item or predicate before writing (reading) that data item, or set of data items defined by a predicate. The transaction is *well-formed* if it has well-formed writes and reads. A transaction has *two-phase writes* (*reads*) if it does not set a new Write (Read) lock on a data item after releasing a Write (Read) lock. A transaction exhibits *two-phase locking* if it does not request any new locks after releasing some lock.

The locks requested by a transaction are of *long duration* if they are held until after the transaction commits or aborts. Otherwise, they are of *short duration*. Typically, short locks are released immediately after the action completes.

If a transaction holds a lock, and another transaction requests a conflicting lock, then the new lock request is not granted until the former transaction's conflicting lock has been released.

The fundamental serialization theorem is that *well-formed two-phase locking* guarantees *serializability* — each history arising under two-phase locking is equivalent to some

| Table 1. ANSI SQL Isolation Levels Defined in terms of the Three Original Phenomena | | | |
|---|---|---|---|
| **Isolation Level** | **P1 (or A1) Dirty Read** | **P2 (or A2) Fuzzy Read** | **P3 (or A3) Phantom** |
| ANSI READ UNCOMMITTED | Possible | Possible | Possible |
| ANSI READ COMMITTED | Not Possible | Possible | Possible |
| ANSI REPEATABLE READ | Not Possible | Not Possible | Possible |
| ANOMALY SERIALIZABLE | Not Possible | Not Possible | Not Possible |

serial history. Conversely, if a transaction is not well-formed or two-phased then, except in degenerate cases, non-serializable execution histories are possible [EGLT].

The [GLPT] paper defined four *degrees of consistency*, attempting to show the equivalence of locking, dependency, and anomaly-based characterizations. The anomaly definitions (see Definition 1) were too vague. The authors continue to get criticism for that aspect of the definitions [GR]. Only the more mathematical definitions in terms of histories and dependency graphs or locking have stood the test of time.

Table 2 defines a number of isolation types in terms of lock *scopes* (items or predicates), *modes* (read or write), and their *durations* (short or long). We believe the isolation levels called Locking READ UNCOMMITTED, Locking READ COMMITTED, Locking REPEATABLE READ, and Locking SERIALIZABLE are the locking definitions *intended* by ANSI SQL Isolation levels — but as shown next they are quite different from those of Table 1. Consequently, it is necessary to differentiate isolation levels defined in terms of locks from the ANSI SQL phenomena-based isolation levels. To make this distinction, the levels in Table 2 are labeled with the "Locking" prefix, as opposed to the "ANSI" prefix of Table 1.

[GLPT] defined Degree 0 consistency to allow both dirty reads and writes: it only required action atomicity. Degrees 1, 2, and 3 correspond to Locking READ UNCOMMITTED, READ COMMITTED, and SERIALIZABLE, respectively. No isolation degree matches the Locking REPEATABLE READ isolation level.

Date and IBM originally used the name "Repeatable Reads" [DAT, DB2] to mean serializable or Locking SERIALIZABLE. This seemed like a more comprehensible name than the [GLPT] term "Degree 3 isolation." The ANSI SQL meaning of REPEATABLE READ is different from Date's original definition, and we feel the terminology is unfortunate. Since anomaly P3 is specifically not ruled out by the ANSI SQL REPEATABLE READ isolation level, it is clear from the definition of P3 that reads are NOT repeatable! We repeat this misuse of the term with Locking REPEATABLE READ in Table 2, in order to parallel the ANSI definition. Similarly, Date coined the term Cursor Stability as a more comprehensible name for Degree 2 isolation augmented with protection from lost cursor updates as explained in Section 4.1 below.

**Definition.** Isolation level L1 is *weaker* than isolation level L2 (or L2 is *stronger* than L1), denoted L1 « L2, if all non-serializable histories that obey the criteria of L2 also satisfy L1 and there is at least one non-serializable history that can occur at level L1 but not at level L2. Two isolation levels L1 and L2 are *equivalent*, denoted L1 == L2, if the sets of non-serializable histories satisfying L1 and L2 are identical. L1 is *no stronger* than L2, denoted L1 «̲ L2 if either L1 « L2 or L1 == L2. Two isolation levels are *incomparable*, denoted L1 »« L2, when each isolation level allows a non-serializable history that is disallowed by the other.

In comparing isolation levels we differentiate them only in terms of the *non-serializable* histories that can occur in one but not the other. Two isolation levels can also differ in terms of the serializable histories they permit, but we say Locking SERIALIZABLE == Serializable even though it is well known that a locking scheduler does not admit all possible Serializable histories. It is possible for an isolation level to be impractical because of disallowing too many serializable histories, but we do not deal with this here.

These definitions yield the following remark.
**Remark 1:** Locking READ UNCOMMITTED
 « Locking READ COMMITTED
 « Locking REPEATABLE READ
 « Locking SERIALIZABLE

| Table 2. Degrees of Consistency and Locking Isolation Levels defined in terms of locks. | | |
|---|---|---|
| **Consistency Level = Locking Isolation Level** | **Read Locks on Data Items and Predicates (the same unless noted)** | **Write Locks on Data Items and Predicates (always the same)** |
| Degree 0 | none required | Well-formed Writes |
| Degree 1 = Locking READ UNCOMMITTED | none required | Well-formed Writes Long duration Write locks |
| Degree 2 = Locking READ COMMITTED | Well-formed Reads Short duration Read locks (both) | Well-formed Writes, Long duration Write locks |
| Cursor Stability (see Section 4.1) | Well-formed Reads Read locks held on current of cursor Short duration Read Predicate locks | Well-formed Writes, Long duration Write locks |
| Locking REPEATABLE READ | Well-formed Reads Long duration data-item Read locks Short duration Read Predicate locks | Well-formed Writes, Long duration Write locks |
| Degree 3 = Locking SERIALIZABLE | Well-formed Reads Long duration Read locks (both) | Well-formed Writes, Long duration Write locks |

In the following section, we'll focus on comparing the ANSI and Locking definitions.

## 3. Analyzing ANSI SQL Isolation Levels

To start on a positive note, the locking isolation levels comply with the ANSI SQL requirements.

**Remark 2**. The locking protocols of Table 2 define locking isolation levels that are at least as strong as the corresponding phenomena-based isolation levels of Table 1. See [OOBBGM] for proof.

Hence, locking isolation levels are at least as isolated as the same-named ANSI levels. Are they more isolated? The answer is yes, even at the lowest level. Locking READ UNCOMMITTED provides long duration write locking to avoid a phenomenon called "Dirty Writes," but ANSI SQL does not exclude this anomalous behavior other than ANSI SERIALIZABLE. Dirty writes are defined as follows:

**P0 (Dirty Write):** Transaction T1 modifies a data item. Another transaction T2 then further modifies that data item before T1 performs a COMMIT or ROLLBACK. If T1 or T2 then performs a ROLLBACK, it is unclear what the correct data value should be. The broad interpretation of this is:

**P0**:  $w1[x]...w2[x]...((c1$ or $a1)$ and $(c2$ or $a2)$ in any order$)$

One reason why Dirty Writes are bad is that they can violate database consistency. Assume there is a constraint between x and y (e.g., x = y), and T1 and T2 each maintain the consistency of the constraint if run alone. However, the constraint can easily be violated if the two transactions write x and y in different orders, which can only happen if there are Dirty writes. For example consider the history w1[x] w2[x] w2[y] c2 w1[y] c1. T1's changes to y and T2's to x both "survive". If T1 writes 1 in both x and y while T2 writes 2, the result will be x=2, y =1 violating x = y.

As discussed in [GLPT, BHG] and elsewhere, automatic transaction rollback is another pressing reason why P0 is important. Without protection from P0, the system can't undo updates by restoring before images. Consider the history: w1[x] w2[x] a1. You don't want to undo w1[x] by restoring its before-image of x, because that would wipe out w2's update. But if you don't restore its before-image, and transaction T2 later aborts, you can't undo w2[x] by restoring *its* before-image either! Even the weakest locking systems hold long duration write locks. Otherwise, their recovery systems would fail. So we conclude Remark 3:

**Remark 3:** ANSI SQL isolation should be modified to require P0 for all isolation levels.

We now argue that a broad interpretation of the three ANSI phenomena is required. Recall the strict interpretations are:

**A1**:  $w1[x]...r2[x]...$(a1 and c2 in either order) (Dirty Read)
**A2**:  $r1[x]...w2[x]...c2...r1[x]...c1$   (Fuzzy or Non-Repeatable Read)
**A3**:  $r1[P]...w2[y$ in $P]...c2....r1[P]...c1$   (Phantom)

By Table 1, histories under READ COMMITTED isolation forbid anomaly A1, REPEATABLE READ isolation forbids anomalies A1 and A2, and SERIALIZABLE isolation forbids anomalies A1, A2, and A3. Consider history H1, involving a $40 transfer between bank balance rows x and y:

H1: r1[x=50]w1[x=10]r2[x=10]r2[y=50]c2 r1[y=50]w1[y=90]c1

H1 is non-serializable, the classical *inconsistent analysis* problem where transaction T1 is transferring a quantity 40 from x to y, maintaining a total balance of 100, but T2 reads an inconsistent state where the total balance is 60. The history H1 does not violate any of the anomalies A1, A2, or A3. In the case of A1, one of the two transactions would have to abort; for A2, a data item would have to be read by the same transaction for a second time; A3 requires a phantom value. None of these things happen in H1. Consider instead taking the broad interpretation of A1, the phenomenon P1:

**P1**:  $w1[x]...r2[x]...((c1$ or $a1)$ and $(c2$ or $a2)$ in any order$)$

H1 indeed violates P1. Thus, we should take the interpretation P1 for what was intended by ANSI rather than A1. The Broad interpretation is the correct one.

Similar arguments show that P2 should be taken as the ANSI intention rather than A2. A history that discriminates these two interpretations is:

H2: r1[x=50]r2[x=50]w2[x=10]r2[y=50]w2[y=90]c2r1[y=90]c1

H2 is non-serializable — it is another inconsistent analysis, where T1 sees a total balance of 140. This time neither transaction reads dirty (i.e. uncommitted) data. Thus P1 is satisfied. Once again, no data item is read twice nor is any relevant predicate evaluation changed. The problem with H2 is that by the time T1 reads y, the value for x is out of date. If T2 were to read x again, it would have been changed; but since T2 doesn't do that, A2 doesn't apply. Replacing A2 with P2, the broader interpretation, solves this problem.

**P2:**  $r1[x]...w2[x]...((c1$ or $a1)$ and $(c2$ or $a2)$ any order$)$

H2 would now be disqualified when w2[x=20] occurs to overwrite r1[x=50]. Finally, consider A3 and history H3:

**A3:**  $r1[P]...w2[y$ in $P]...c2...r1[P]...c1$   (Phantom)

H3:  r1[P] w2[insert y to P] r2[z] w2[z] c2 r1[z] c1

Here T1 performs a `<search condition>` to find the list of active employees. Then T2 performs an insert of a new active employee and then updates z, the count of employees in the company. Following this, T1 reads the count of active employees as a check and sees a discrepancy. This history is clearly not serializable, but is allowed by A3 since no predicate is evaluated twice. Again, the Broad interpretation solves the problem.

**P3:** r1[P]...w2[y in P]...((c1 or a1) and (c2 or a2) any order)

If P3 is forbidden, history H3 is invalid. This is clearly what ANSI intended. The foregoing discussion demonstrates the following results.

**Remark 4**. Strict interpretations A1, A2, and A3 have unintended weaknesses. The correct interpretations are the Broad ones. We assume in what follows that ANSI meant to define P1, P2, and P3.

**Remark 5**. ANSI SQL isolation phenomena are incomplete. There are a number of anomalies that still can arise. New phenomena must be defined to complete the definition of locking. Also, P3 must be restated. In the following definitions, we drop references to (c2 or a2) that do not restrict histories.

**P0**: w1[x]...w2[x]...(c1 or a1)      (Dirty Write)
**P1**: w1[x]...r2[x]...(c1 or a1)      (Dirty Read)
**P2**: r1[x]...w2[x]...(c1 or a1)      (Fuzzy or Non-Repeatable Read)
**P3**: r1[P]...w2[y in P]...(c1 or a1)     (Phantom)

One important note is that ANSI SQL P3 only prohibits inserts (and updates, according to some interpretations) to a predicate whereas the definition of P3 above prohibits *any* write satisfying the predicate once the predicate has been read — the write could be an insert, update, or delete.

The definition of proposed ANSI isolation levels in terms of these phenomena is given in Table 3.
For single version histories, it turns out that the P0, P1, P2, P3 phenomena are disguised versions of locking. For example, prohibiting P0 precludes a second transaction writing an item after the first transaction has written it, equivalent to saying that long-term Write locks are held on data items (and predicates). Thus Dirty Writes are impossible at all levels. Similarly, prohibiting P1 is equivalent to having well-formed reads on data items. Prohibiting P2 means long-term Read locks on data items. Finally, Prohibiting P3 means long-term Read predicate locks. Thus the isolation levels of Table 3 defined by these phenomena provide the same behavior as the Locking isolation levels of Table 2.

**Remark 6**. The locking isolation levels of Table 2 and the phenomenological definitions of Table 3 are equivalent. Put another way, P0, P1, P2, and P3 are disguised redefinition's of locking behavior.

In what follows, we will refer to the isolation levels listed in Table 3 by the names in Table 3, equivalent to the Locking versions of these isolation levels of Table 2. When we refer to ANSI READ UNCOMMITTED, ANSI READ COMMITTED, ANSI REPEATABLE READ, and ANOMALY SERIALIZABLE, we are referring to the ANSI definition of Table 1 (inadequate, since it did not include P0).

The next section shows that a number of commercially available isolation implementations provide isolation levels that fall between READ COMMITTED and REPEATABLE READ. To achieve meaningful isolation levels that distinguish these implementations, we will assume P0 and P1 as a basis and then add distinguishing new phenomena.

## 4. Other Isolation Types

### 4.1 Cursor Stability

Cursor Stability is designed to prevent the lost update phenomenon.

**P4 (Lost Update)**: The lost update anomaly occurs when transaction T1 reads a data item and then T2 updates the data item (possibly based on a previous read), then T1 (based on its earlier read value) updates the data item and commits. In terms of histories, this is:

**P4:** r1[x]...w2[x]...w1[x]...c1      (Lost Update)

The problem, as illustrated in history H4, is that even if T2 commits, T2's update will be lost.

H4:   r1[x=100] r2[x=100] w2[x=120] c2 w1[x=130] c1

| Table 3. ANSI SQL Isolation Levels Defined in terms of the four phenomena ||||| 
|---|---|---|---|---|
| **Isolation Level** | **P0** Dirty Write | **P1** Dirty Read | **P2** Fuzzy Read | **P3** Phantom |
| **READ UNCOMMITTED** | Not Possible | Possible | Possible | Possible |
| **READ COMMITTED** | Not Possible | Not Possible | Possible | Possible |
| **REPEATABLE READ** | Not Possible | Not Possible | Not Possible | Possible |
| **SERIALIZABLE** | Not Possible | Not Possible | Not Possible | Not Possible |

The final value of x contains only the increment of 30 added by T1. P4 is possible at the READ COMMITTED isolation level, since H4 is allowed when forbidding P0 (a commit of the transaction performing the first write action precedes the second write) or P1 (which would require a read after a write). However, forbidding P2 also precludes P4, since w2[x] comes after r1[x] and before T1 commits or aborts. Therefore the anomaly P4 is useful in distinguishing isolation levels intermediate in strength between READ COMMITTED and REPEATABLE READ.

The *Cursor Stability* isolation level extends READ COMMITTED locking behavior for SQL cursors by adding a new read action for FETCH from a cursor and requiring that a lock be held on the current item of the cursor. The lock is held until the cursor moves or is closed, possibly by a commit. Naturally, the Fetching transaction can update the row, and in that case a write lock will be held on the row until the transaction commits, even after the cursor moves on with a subsequent Fetch. The notation is extended to include, rc, meaning read cursor, and wc, meaning write the current record of the cursor. A rc1[x] and a later wc1[x] precludes an intervening w2[x]. Phenomenon P4, renamed P4C, is prevented in this case.

  P4C:    rc1[x]...w2[x]...w1[x]...c1          (Lost Update)

**Remark 7:**
READ COMMITTED « Cursor Stability « REPEATABLE READ

Cursor Stability is widely implemented by SQL systems to prevent lost updates for rows read via a cursor. READ COMMITTED, in some systems, is actually the stronger Cursor Stability. The ANSI standard allows this.

The technique of putting a cursor on an item to hold its value stable can be used for multiple items, at the cost of using multiple cursors. Thus the programmer can parlay Cursor Stability to effective Locking REPEATABLE READ isolation for any transaction accessing a small, fixed number of data items. However this method is inconvenient and not at all general. Thus there are always histories fitting the P4 (and of course the more general P2) phenomenon that are not precluded by Cursor Stability.

## 4.2 Snapshot Isolation

These discussions naturally suggest an isolation level, called *Snapshot Isolation*, in which each transaction reads reads data from a snapshot of the (committed) data as of the time the transaction started, called its *Start-Timestamp*. This time may be any time before the transaction's first Read. A transaction running in Snapshot Isolation is never blocked attempting a read as long as the snapshot data from its Start-Timestamp can be maintained. The transaction's writes (updates, inserts, and deletes) will also be reflected in this snapshot, to be read again if the transaction accesses (i.e., reads or updates) the data a second time. Updates by other transactions active after the transaction Start-Timestamp are invisible to the transaction.

Snapshot Isolation is a type of multiversion concurrency control. It extends the Multiversion Mixed Method described in [BHG], which allowed snapshot reads by read-only transactions.

When the transaction T1 is ready to commit, it gets a *Commit-Timestamp,* which is larger than any existing Start-Timestamp or Commit-Timestamp. The transaction successfully commits only if no other transaction T2 with a Commit-Timestamp in T1's *execution interval* [*Start-Timestamp*, *Commit-Timestamp*] wrote data that T1 also wrote. Otherwise, T1 will abort. This feature, called *First-committer-wins* prevents lost updates (phenomenon P4). When T1 commits, its changes become visible to all transactions whose Start-Timestamps are larger than T1's Commit-Timestamp.

Snapshot Isolation is a multi-version (MV) method, so single-valued (SV) histories do not properly reflect the temporal action sequences. At any time, each data item might have multiple versions, created by active and committed transactions. Reads by a transaction must choose the appropriate version. Consider history H1 at the beginning of Section 3, which shows the need for P1 in a single valued execution. Under Snapshot Isolation, the same sequence of actions would lead to the multi-valued history:

H1.SI:   r1[x0=50] w1[x1=10] r2[x0=50] r2[y0=50] c2
         r1[y0=50] w1[y1=90] c1

H1.SI has the dataflows of a serializable execution. In [OOBBGM], we show that all Snapshot Isolation histories can be mapped to single-valued histories while preserving dataflow dependencies (the MV histories are said to be View Equivalent with the SV histories, an approach covered in [BHG], Chapter 5). For example the MV history H1.SI would map to the serializable SV history:

H1.SI.SV:   r1[x=50] r1[y=50] r2[x=50] r2[y=50] c2
            w1[x=10] w1[y=90] c1

Mapping of MV histories to SV histories is the only rigorous touchstone needed to place Snapshot Isolation in the Isolation Hierarchy.

Snapshot Isolation is non-serializable because a transaction's Reads come at one instant and the Writes at another. For example, consider the single-value history:

H5:   r1[x=50] r1[y=50] r2[x=50] r2[y=50] w1[y=-40] w2[x=-40] c1 c2

H5 is non-serializable and has the same inter-transactional dataflows as could occur under Snapshot Isolation (there is no choice of versions read by the transactions). Here we assume that each transaction that writes a new value for x and y is expected to maintain the constraint that x + y should be positive, and while T1 and T2 both act properly in isolation, the constraint fails to hold in H5.

*Constraint violation* is a generic and important type of concurrency anomaly. Individual databases satisfy constraints over multiple data items (e.g., uniqueness of keys, referential integrity, replication of rows in two tables, etc.). Together they form the database invariant constraint predicate, *C(DB)*. The invariant is *TRUE* if the database state *DB* is consistent with the constraints and is *FALSE* otherwise. Transactions must preserve the constraint predicate to maintain consistency: if the database is consistent when the transaction starts, the database will be consistent when the transaction commits. If a transaction reads a database state that violates the constraint predicate, then the transaction suffers from a constraint violation concurrency anomaly. Such constraint violations are called *inconsistent analysis* in [DAT].

**A5 (Data Item Constraint Violation)**. Suppose *C()* is a database constraint between two data items x and y in the database. Here are two anomalies arising from constraint violation.

**A5A Read Skew** Suppose transaction T1 reads x, and then a second transaction T2 updates x and y to new values and commits. If now T1 reads y, it may see an inconsistent state, and therefore produce an inconsistent state as output. In terms of histories, we have the anomaly:

  A5A:   r1[x]...w2[x]...w2[y]...c2...r1[y]...(c1 or a1)
                                                              (Read Skew)

**A5B Write Skew** Suppose T1 reads x and y, which are consistent with *C()*, and then a T2 reads x and y, writes x, and commits. Then T1 writes y. If there were a constraint between x and y, it might be violated. In terms of histories:
  A5B:   r1[x]...r2[y]...w1[y]...w2[x]...(c1 and c2 occur)
                                                              (Write Skew)

Fuzzy Reads (P2) is a degenerate form of Read Skew where x=y. More typically, a transaction reads two different but related items (e.g., referential integrity). Write Skew (A5B) could arise from a constraint at a bank, where account balances are allowed to go negative as long as the sum of commonly held balances remains non-negative, with an anomaly arising as in history H5.

Clearly neither A5A nor A5B could arise in histories where P2 is precluded, since both A5A and A5B have T2 write a data item that has been previously read by an uncommitted T1. Thus, phenomena A5A and A5B are only useful for distinguishing isolation levels that are below REPEATABLE READ in strength.

The ANSI SQL definition of REPEATABLE READ, in its strict interpretation, captures a degenerate form of row constraints, but misses the general concept. To be specific, Locking REPEATABLE READ of Table 2 provides protection from Row Constraint Violations but the ANSI SQL definition of Table 1, forbidding anomalies A1 and A2, does not.

Returning now to Snapshot Isolation, it is surprisingly strong, even stronger than READ COMMITTED.

**Remark 8**. READ COMMITTED « Snapshot Isolation

**Proof**. In Snapshot Isolation, first-committer-wins precludes P0 (dirty writes), and the timestamp mechanism prevents P1 (dirty reads), so Snapshot Isolation is no weaker than READ COMMITTED. In addition, A5A is possible under READ COMMITTED, but not under the Snapshot Isolation timestamp mechanism. Therefore READ COMMITTED « Snapshot Isolation.

Note that it is difficult to picture how Snapshot Isolation histories can disobey phenomenon P2 in the single-valued interpretation. Anomaly A2 cannot occur, since a transaction under Snapshot Isolation will read the same value of a data item even after a temporally intervening update by another transaction. However, Write Skew (A5B) obviously can occur in a Snapshot Isolation history (e.g., H5), and in the Single Valued history interpretation we've been reasoning about, forbidding P2 also precludes A5B. Therefore Snapshot Isolation admits history anomalies that REPEATABLE READ does not.

Snapshot Isolation cannot experience the A3 anomaly. A transaction rereading a predicate after an update by another will always see the same old set of data items. But the REPEATABLE READ isolation level can experience A3 anomalies. Snapshot Isolation histories prohibit histories with anomaly A3, but allow A5B, while REPEATABLE READ does the opposite. Therefore:

**Remark 9**. REPEATABLE READ »« Snapshot Isolation.

However, Snapshot Isolation does not preclude P3. Consider a constraint that says a set of job tasks determined by a predicate cannot have a sum of hours greater than 8. T1 reads this predicate, determines the sum is only 7 hours and adds a new task of 1 hour duration, while a concurrent transaction T2 does the same thing. Since the two transactions are inserting different data items (and different index entries as well, if any), this scenario is not precluded by First-Committer-Wins and can occur in Snapshot Isolation. But in any equivalent serial history, the phenomenon P3 would arise under this scenario.

Perhaps most remarkable of all, Snapshot Isolation has no phantoms (in the strict sense of the ANSI definitions A3). Each transaction never sees the updates of concurrent transactions. So, one can state the following surprising result (recall that section Table 1 defined ANOMALY SERIALIZABLE as ANSI SQL definition of SERIALIZABLE) without the extra restriction in Subclause 4.28 in [ANSI]:

**Remark 10**. Snapshot Isolation histories preclude anomalies A1, A2 and A3. Therefore, in the anomaly interpretation of ANOMALY SERIALIZABLE of Table 1:
ANOMALY SERIALIZABLE « SNAPSHOT ISOLATION.

Snapshot Isolation gives the freedom to run transactions with very old timestamps, thereby allowing them to do time travel — taking a historical perspective of the database — while never blocking or being blocked by writes. Of course, update transactions with very old timestamps would abort if they tried to update any data item that had been updated by more recent transactions.

Snapshot Isolation admits a simple implementation modeled on the work of Reed [REE]. There are several commercial implementations of such multi-version databases. Borland's InterBase 4 [THA] and the engine underlying Microsoft's Exchange System both provide Snapshot Isolation with the First-committer-wins feature. First-committer-wins requires the system to remember all updates (write locks) belonging to any transaction that commits after the Start-Timestamp of each active transaction. It aborts the transaction if its updates conflict with remembered updates by others.

Snapshot Isolation's "optimistic" approach to concurrency control has a clear concurrency advantage for read-only transactions, but its benefits for update transactions is still debated. It probably isn't good for long-running update transactions competing with high-contention short transactions, since the long-running transactions are unlikely to be the first writer of everything they write, and so will probably be aborted. (Note that this scenario would cause a real problem in locking implementations as well, and if the solution is to not allow long-running update transactions that would hold up short transaction locks, Snapshot Isolation would also be acceptable.) Certainly in cases where short update transactions conflict minimally and long-running transactions are likely to be read only, Snapshot Isolation should give good results. In regimes where there is high contention among transactions of comparable length, Snapshot Isolation offers a classical optimistic approach, and there are differences of opinion as to the value of this.

### 4.3 Other Multi-Version Systems

There are other models of multi-versio99ning. Some commercial products maintain versions of objects but restrict Snapshot Isolation to read-only transactions (e.g., SQL-92, Rdb, and `SET TRANSACTION READ ONLY` in some other databases [MS, HOB, ORA]; Postgres and Illustra [STO, ILL] maintain such versions long-term and provide time-travel queries). Others allow update transactions but do not provide first-committer-wins protection (e.g., Oracle Read Consistency isolation [ORA]).

Oracle *Read Consistency* isolation gives each SQL statement the most recent committed database value at the time the statement began. It is as if the start-timestamp of the transaction is advanced at each SQL statement. The members of a cursor set are as of the time of the Open Cursor. The underlying mechanism recomputes the appropriate version of the row as of the statement timestamp. Row inserts, updates, and deletes are covered by Write locks to give a first-writer-wins rather than a first-committer-wins policy. Read Consistency is stronger than READ COMMITTED (it disallows cursor lost updates (P4C)) but allows non-repeatable reads (P3), general lost updates (P4), and read skew (A5A). Snapshot Isolation does not permit P4 or A5A.

If one looks carefully at the SQL standard, it defines each statement as atomic. It has a serializable sub-transaction (or timestamp) at the start of each statement. One can imagine a hierarchy of isolation levels defined by assigning timestamps to statements in interesting ways (e.g., in Oracle, a cursor fetch has the timestamp of the cursor open).

## 5. Summary and Conclusions

In summary, there are serious problems with the original ANSI SQL definition of isolation levels (as explained in Section 3). The English language definitions are ambiguous and incomplete. Dirty Writes (P0) are not precluded. Remark 5 is our recommendation for cleaning up the ANSI Isolation levels to equate to the locking isolation levels of [GLPT].

ANSI SQL intended to define REPEATABLE READ isolation to exclude all anomalies except Phantom. The anomaly definition of Table 1 does not achieve this goal, but the locking definition of Table 2 does. ANSI's choice of the term Repeatable Read is doubly unfortunate: (1) repeatable reads do not give repeatable results, and (2) the industry had already used the term to mean exactly that: repeatable reads mean serializable in several products. We recommend that another term be found for this.

| | **P0** Dirty Write | **P1** Dirty Read | **P4C** Cursor Lost Update | **P4** Lost Update | **P2** Fuzzy Read | **P3** Phantom | **A5A** Read Skew | **A5B** Write Skew |
|---|---|---|---|---|---|---|---|---|
| **Isolation level** | | | | | | | | |
| READ UNCOMMITTED == Degree 1 | Not Possible | Possible | Possible | Possible | Possible | Possible | Possible | Possible |
| READ COMMITTED == Degree 2 | Not Possible | Not Possible | Possible | Possible | Possible | Possible | Possible | Possible |
| Cursor Stability | Not Possible | Not Possible | Not Possible | Sometimes Possible | Sometimes Possible | Possible | Possible | Sometimes Possible |
| REPEATABLE READ | Not Possible | Not Possible | Not Possible | Not Possible | Not Possible | Possible | Not Possible | Not Possible |
| Snapshot | Not Possible | Not Possible | Not Possible | Not Possible | Not Possible | Sometimes Possible | Not Possible | Possible |
| ANSI SQL SERIALIZABLE == Degree 3 == Repeatable Read Date, IBM, Tandem, ... | Not Possible | Not Possible | Not Possible | Not Possible | Not Possible | Not Possible | Not Possible | Not Possible |

**Table 4**. Isolation Types Characterized by Possible Anomalies Allowed.

A number of commercially-popular isolation levels, falling between the REPEATABLE READ and SERIALIZABLE levels of Table 3 in strength, have been characterized with some new phenomena and anomalies in Section 4. All the isolation levels named here have been characterized as shown in Figure 2 and Table 4. Isolation levels at higher levels in Figure 2 are higher in strength (see the Definition at the beginning of Section 4.1) and the connecting lines are labeled with the phenomena and anomalies that differentiate them.

On a positive note, reduced isolation levels for multi-version systems have never been characterized before — despite being implemented in several products. Many applications avoid lock contention by using Cursor Stability or Oracle's Read Consistency isolation. Such applications will find Snapshot Isolation better behaved than either: it avoids the lost update anomaly, some phantom anomalies (e.g., the one defined by ANSI SQL), it never blocks read-only transactions, and readers do not block updates.

## Acknowledgments.

We thank Chris Larson of Microsoft, Alan Reiter who pointed out a number of the newer anomalies in Snapshot Isolation, Franco Putzolu and Anil Nori of Oracle, Mike Ubell of Illustra and the anonymous SIGMOD referees for valuable suggestions that improved this paper. Sushil Jajodia, V. Atluri, and E. Bertino generously provided us with an early draft of their related work [ABJ] on reduced isolation levels for multi-valued histories.

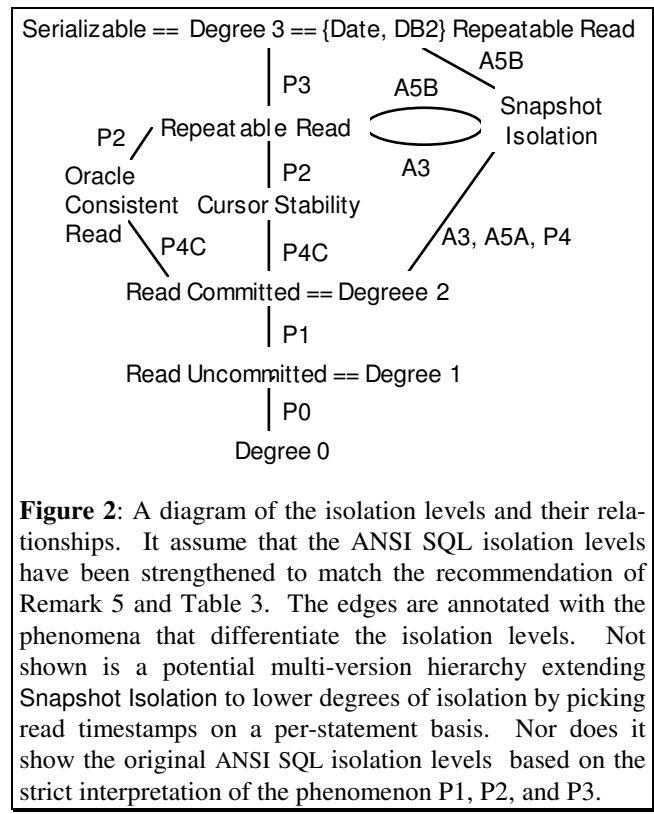

**Figure 2**: A diagram of the isolation levels and their relationships. It assume that the ANSI SQL isolation levels have been strengthened to match the recommendation of Remark 5 and Table 3. The edges are annotated with the phenomena that differentiate the isolation levels. Not shown is a potential multi-version hierarchy extending Snapshot Isolation to lower degrees of isolation by picking read timestamps on a per-statement basis. Nor does it show the original ANSI SQL isolation levels based on the strict interpretation of the phenomenon P1, P2, and P3.